\documentclass[twocolumn,prl,showpacs,amsfonts,superscriptaddress,floatfix,nofootinbib,10pt]{revtex4-1}
\usepackage{amsmath}
\usepackage{graphicx}
\usepackage{footmisc}
\pdfoutput=1
\usepackage{color}

\begin{document}
\title {Edge modes in the  Hofstadter model of interacting electrons}
\author{Igor N. Karnaukhov}
\affiliation{G.V. Kurdyumov Institute for Metal Physics, 36 Vernadsky Boulevard, 03142 Kiev, Ukraine}
\begin{abstract}
We provide a detailed analysis of a realization of chiral gapless edge modes in the framework of the Hofstadter model of interacting electrons. In a transverse homogeneous magnetic field and a rational magnetic flux through an unit cell the fermion spectrum splits into topological subbands with well-defined Chern numbers, contains gapless edge modes in the gaps. It is shown that the behavior of gapless edge modes is described within the framework of the Kitaev chain where the tunneling of Majorana fermions is determined by effective hopping of Majorana fermions between chains. The proposed approach makes it possible to study the fermion spectrum in the case of an irrational flux, to calculate the Hall conductance of subbands that form a fine structure of the spectrum. In the case of a rational flux and a strong on-site Hubbard interaction $U$, $ U >4 \Delta $ ($ \Delta $ is a gap), the topological state of the system, which is determined by the corresponding Chern number and chiral gapless edge modes, collapses. When the magnitude of the on-site Hubbard interaction changes, at the point $ U = 4 \Delta $ a topological phase transition is realized, i.e., there are changes in the Chern numbers of two subbands due to their degeneration.
\end{abstract}
\maketitle

\section{Introduction}

Exotic phases in the condensed matter physics may arise in topologically nontrivial superconductors, insulators. Topological states are described in the framework of the band theory, while the topological invariant characterizes the class of a topological insulator or a topological superconductor \cite{0,9,10}. A ground state of the Chern systems is characterized by the Chern number, it is an integer topological number which is well defined in insulator (superconductor) state for the band isolated from all other bands \cite{H,K1,K2}. A non-trivial topology of the ground state provides the quantization of the Hall conductance, which can be interpreted as the  Chern number. The value of the Chern number for a given occupied band is far from obvious without numerical calculations, so another approach to the study of nontrivial topological states of the system is the existence of chiral gapless edge modes.

In the case of a rational magnetic flux  per unit cell, the Chern insulator states are realized in the Hofstadter model.  All possible energies of fermions  resemble a butterfly and is referred as the Hofstadter butterfly \cite{Hof}. The competition of two periods (the lattice and the magnetic sublattices) lies at the heart of the butterfly graph. The Chern number $\sigma_\gamma$ of filled $\gamma$ bands satisfies the Diophantine equation $p \sigma_\gamma =  q \cdot s + \gamma$  \cite{D1,D2} (where $s$ is an integer). Equation follows from magnetic translational invariance of the system in the case of a rational magnetic flux through the unit cell $\phi$, that is determined in the quantum flux unit $\phi= p/q$ ($p,q$ are coprime integers).  The phase diagram representing the values of the Chern numbers as a function of a magnetic flux is known as the colored Hofstadter butterfly \cite{4},\cite{5}.

Although the Chern number was originally determined in the momentum space; several generalizations have been constructed for quasiperiodic systems \cite {Ch1}, amorphous topological insulators \cite {Ch2}. In the case of an irrational flux the Chern number is not clearly defined, since the Brillouin zone is not defined (the Chern number is not determined as an integer for quasiperiodic systems \cite{Ch1}, the bulk-edge correspondence is broken). In the case of an irrational magnetic flux it difficult to calculate the topological numbers using the system-dependent approach based on successive rational approximants of the irrational flux.  Most likely the topological order is determined by the total number of gapless edge modes.

Initially, the Chern insulators were solved in the framework of noninteracting fermion models with periodic and open boundary conditions. Most of the topological states found in condensed matter systems belong to different classes of noninteracting topological insulators. It is expected that weak interactions will not significantly change the stability of topological states. The behavior of topological phases is much more complicated, however,
once interactions are taken into account.
The interaction between particles can lead to a transition from topological to topological trivial phase, as a result of which a classification of noninteracting  fermion systems can collapse in a real system.
The  slave-particles  mean-field theory \cite{11,12,13}, the cellular dynamical mean-field theory \cite{14}, quantum Monte Carlo simulations \cite{15,16} and other methods of analytical or numerical calculations of interacting fermion system do not always work in low dimensional and 2D, in particular, systems with interaction.
In \cite{17,18,19} the authors have explored the topological order in 1D Majorana fermion models in context of the Kitaev p-wave chain model with interaction, which exhibits $\mathbf{Z}_2$ topological order.
Strong interaction in the 1D electron systems destroy the superconducting gap that stabilizes the Majorana edge states. In contrast to 2D fermion models, some 1D models with interaction are exactly solvable \cite{KOR, Kar}, which allows us to investigate the effect of strong interaction on the phase state.
The aim of the work is to make a nontrivial step in the study of topological systems taking into account the interaction, namely, to obtain a stability criterion for the state of a 2D topological insulator taking into account the on-site Hubbard interaction between electrons

The question arises: how important is the interaction between fermions for the stability of chiral gapless edge modes, hence, and the stability of the topological state? Within the Hofstadter model, as the model of the 2D topological Chern insulator, in which an external magnetic field breaks a time reversal symmetry, we found the answer to this question. Using the exact solution of the fermion chain model \cite{MN}, we determine the stability of topological state in the Hofstadter model of interacting electrons.
We show that the interaction leads to the topological phase transition at a critical value of the on-site Hubbard interaction, the point of the phase transition separates the state with gapless Majorana edge modes and topological trivial phase. We propose the effective Hamiltonian which describes the low energy fermion states near and in the  gaps, that has exact solution for an arbitrary strength of the on-site Hubbard interaction.
\section{Model Hamiltonian}

We consider the 2D Chern insulator defined on a square lattice within the Hofstadter model \cite{Hof}.
In the case of a rational magnetic flux through a unit cell, the Hofstadter model is reduced to one-particle model of spinless fermions with the q-magnetic cell, which leads to commensurability of magnetic and lattice scales \cite{Har}. In the presence of a transverse homogeneous magnetic field $H{\textbf{e}}_z$ the model Hamiltonian has a well-known form \cite{Hof}
\begin{eqnarray}
 {\cal H}_0=  \sum_{n,j} [t^x(n,j) a^\dagger_{n,j} a_{n+1,j} + t^y(n,j)a^\dagger_{n,j}  a_{n,j+1} + H.c.]
    \label{eq:H0}
\end{eqnarray}
where $a^\dagger_{n,j} $ and $a_{n,j}$ are the fermion operators determined on a square lattice with sites $\{n,j\}$. The one-particle Hamiltonian (1) describes the nearest-neighbor hoppings of spinless fermions with different amplitudes along the $x$-direction $t^x(n,j)=t$ and along the $y$-direction  $t^y (n,j)=\exp[2i \pi (n-1)\phi]$. A magnetic flux through the unit cell $\phi = \frac{H }{ \Phi_0}$ is determined in the quantum flux unit ${\Phi_0=h/e}$, a homogeneous field~$H$ is represented by its vector potential $\textbf{A} = H x \textbf{e}_y$. The value $t$ changes in the interval $0\leq t\leq 1$, the limiting cases of weak  $t=0$ and strong $t=1$ hopping integrals correspond to isolated y-chains and the original Hofstadter model \cite{Hof}, respectively. We consider the 2D fermion system of size $N \times N$ in a hollow cylindrical geometry with open boundary conditions (a cylinder axis along the $x$-direction and the boundaries along the y-direction). For an arbitrary magnetic flux the model is reduced to a quasiperiodic 1D system of length N.

\section{Topological structure of the spectrum}

\subsection{A rational flux $\phi =\frac{p}{q}$}

The behavior of the fermion spectrum of the Hamiltonian (1) in a transverse homogeneous magnetic field is studied in \cite{IK}. We will use this approach to study stability of topological states of interacting fermion systems.
First of all we consider the case $p=1, q=3$, the one-particle spectrum splits into three isolated subbands (see in {fig:1}). Three subbands, each of which is isolated from all other subbands, are topological with the following Chern numbers $\{1,-2,1\}$. The topological subbands are connected by the chiral edge modes, the total number of edges modes in the gap $\Delta_\gamma$  with allowance for their chirality, determines the Chern number of $\gamma$-filled subbands $\sigma_\gamma=\sum_{\alpha =1}^\gamma C_\alpha$.
For a detailed study of the peculiarities of the fermion spectrum  we consider an anisotropic variant of the Hofstadter model with an arbitrary $t$. The values of gaps are equal to $\Delta = 2 |t|-0(t^2) $ at $\epsilon= \pm 1$ (see in {fig:1}).

In the case $t<<1$ the fermions form  the chains along the \emph{y}-direction with weak tunneling of fermions between the chains. The fermion states are determined by the Bloch function, with a wave vector $k$ directed along the y-chains. The fermion states in the gaps are calculated for a fixed $k$ corresponding to energy of the middle of the gap $\varepsilon$. In the $t\to 0$ limit the energies of fermions in the chains are shifted by $k_0=\frac{2\pi}{q}$ or $\frac{2\pi}{3}$ for $q=3$, intersect at the points $k_p= \frac{(2p-1) \pi}{3}$, $\epsilon=\pm 1$ ($p=1,2,3$) (see in {fig:1}c)).
We can define the tunneling of fermions between the $y$-chains as hoppings of fermions between nearest-neighbor lattice sites along the $x$-chain $\tau \sum_{n=1}^{N-1} a_{n}^\dagger a_{n+1}+H.c.$, where $\tau$ is the effective hopping integral and $n$ is the number of the y-chain. Using the presentation for the Majorana fermions $\chi_{n}=a_{n}+ a_{n}^\dagger$ and $\gamma_{n}=\frac{ a_{n}- a_{n}^\dagger}{i}$ we redefine this term in the following form $i\frac{ \tau}{2} \sum_{n=1}^{N-1}( \chi_{n} \gamma_{n+1} -\gamma_{n} \chi_{n+1})$.  Majorana operators $\gamma (j)$  are defined by the algebra $\{\gamma (j),\gamma (i)\}=2 \delta_{j,i}$ and $\gamma (j)=\gamma^\dagger (j)$.
We take into account the tunneling between the fermions at the points of the energy crossings, for which the conservation of momentum and energy is automatically realized, at the same time the chirality of fermions is not conserved. There is the tunneling of fermions between states with different signs of velocities and a certain chirality.

At the energies $\pm 1$ the gaps in the fermion spectrum are determined by tunneling of fermions between corresponding nearest-neighbor $y-$chains, as shown in {fig:1}b),c). At $\varepsilon=1$ for one q-cell there are the following processes of tunneling of fermions: the fermions of the second chain with positive $\frac{\sqrt 3}{2}$ at $-k_1$ and negative $-\frac{\sqrt 3}{2}$ at $k_1$ velocities (black line in {fig:1}c)) tunnel to the first chain at $k_1$ in the state with a positive velocity $\frac{\sqrt 3}{2}$ (red line) and into the third chain at $-k_1$  in the state with negative velocity $-\frac{\sqrt 3}{2}$ (blue line). The result of the tunneling of fermions does not depend on the wave vector, since the tunneling of particles occurs between states with the same wave vectors. Let us write the tunneling processes between the fermions located at  n=2 y-chain and  n=1 $\tau \gamma_2 \chi_1 $ and n=3  $\tau \gamma_3 \chi_2 $ chains.  Majorana operators connect the states of fermions with energies $\pm \varepsilon $  with states of particles and holes.
The tunneling process of fermion between the $y$-chains in the gaps occurs with a change of the chirality of fermions, it is described as $\chi_{n-1} \gamma_{n}$, $\chi_{n} \gamma_{n+1}$ ... (at  $\varepsilon=1$) or  $\chi_{n} \gamma_{n-1}$, $\chi_{n+1} \gamma_{n}$...(at $\varepsilon=-1$) movement of fermions along the $x$-chain. We note, that these processes of tunneling are determined by the conservation of momentum and energy for a fixed energy $\varepsilon$.
In contrast to traditional tunneling of fermions, in which result of tunneling does not depend on the direction of the fermion wave vector, in this case tunneling of fermions with different velocities are realized at different energies $\pm \varepsilon$. We define the effective Hamiltonian, which takes into account the low energy excitations of Majorana fermions in the gap near the energy $\epsilon$
\begin{eqnarray}
&&{\cal H}_{eff}(\epsilon=1) = i\frac{\tau}{2}  \sum_{n=1}^{N-1} \chi_{n} \gamma_{n+1}, \nonumber \\
&&{\cal H}_{eff}(\epsilon=-1) = i\frac{\tau}{2}  \sum_{n=1}^{N-1} \chi_{n+1} \gamma_{n}.
\label{eq:H1}
\end{eqnarray}
In the case of a rational flux $N=q N_c$ ($N_c$ is the number of the q-unit cells), we can consider the states of fermions of one q-cell or $N_c$ cells, taking into account the tunneling of fermions belonging the nearest-neighbor cells.

\begin{figure}[tp]
     \centering{\leavevmode}
    \begin{minipage}[h]{.75\linewidth}
\center{\includegraphics[width=\linewidth]{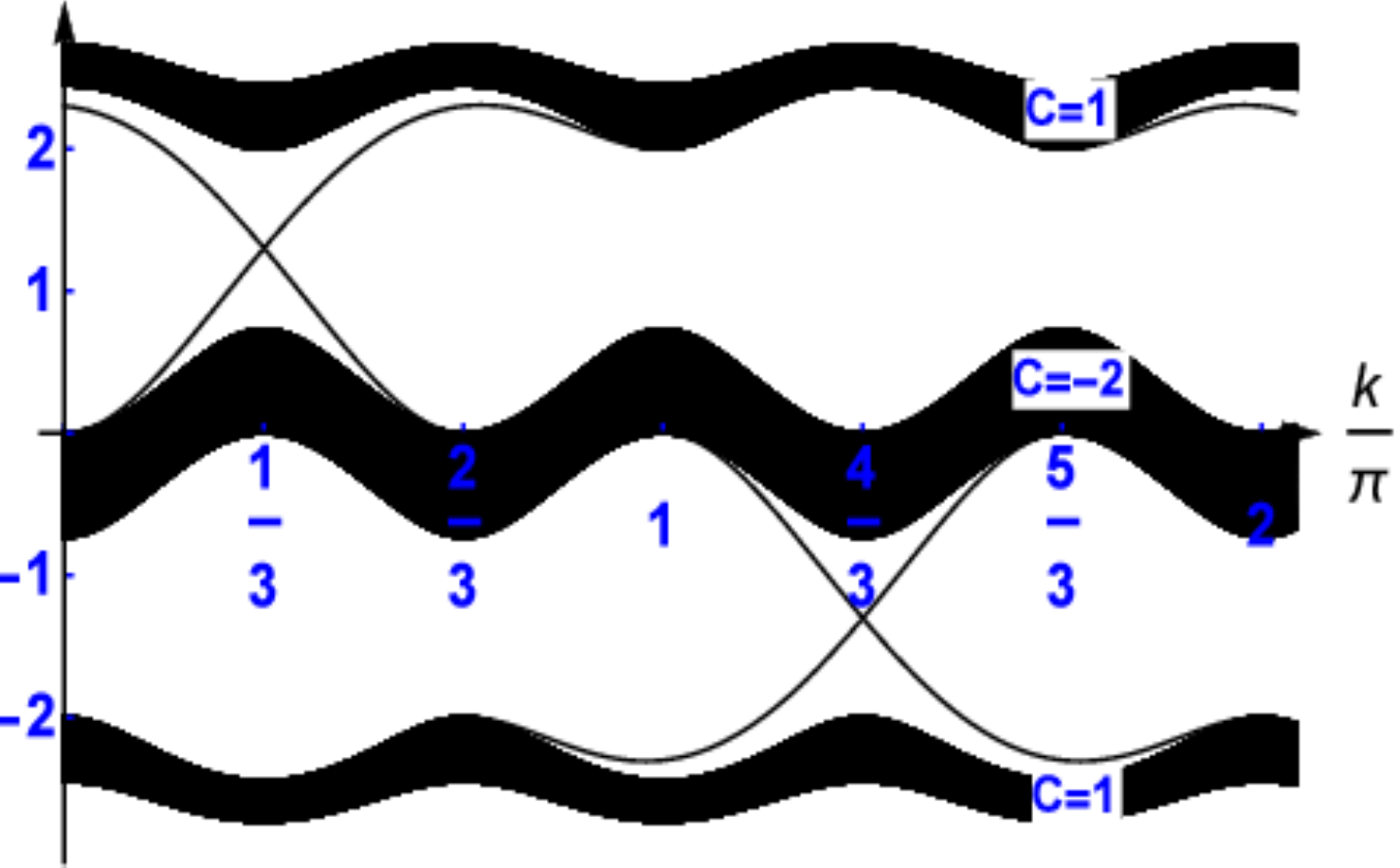} a)}
\end{minipage}
 \centering{\leavevmode}
    \begin{minipage}[h]{.75\linewidth}
\center{\includegraphics[width=\linewidth]{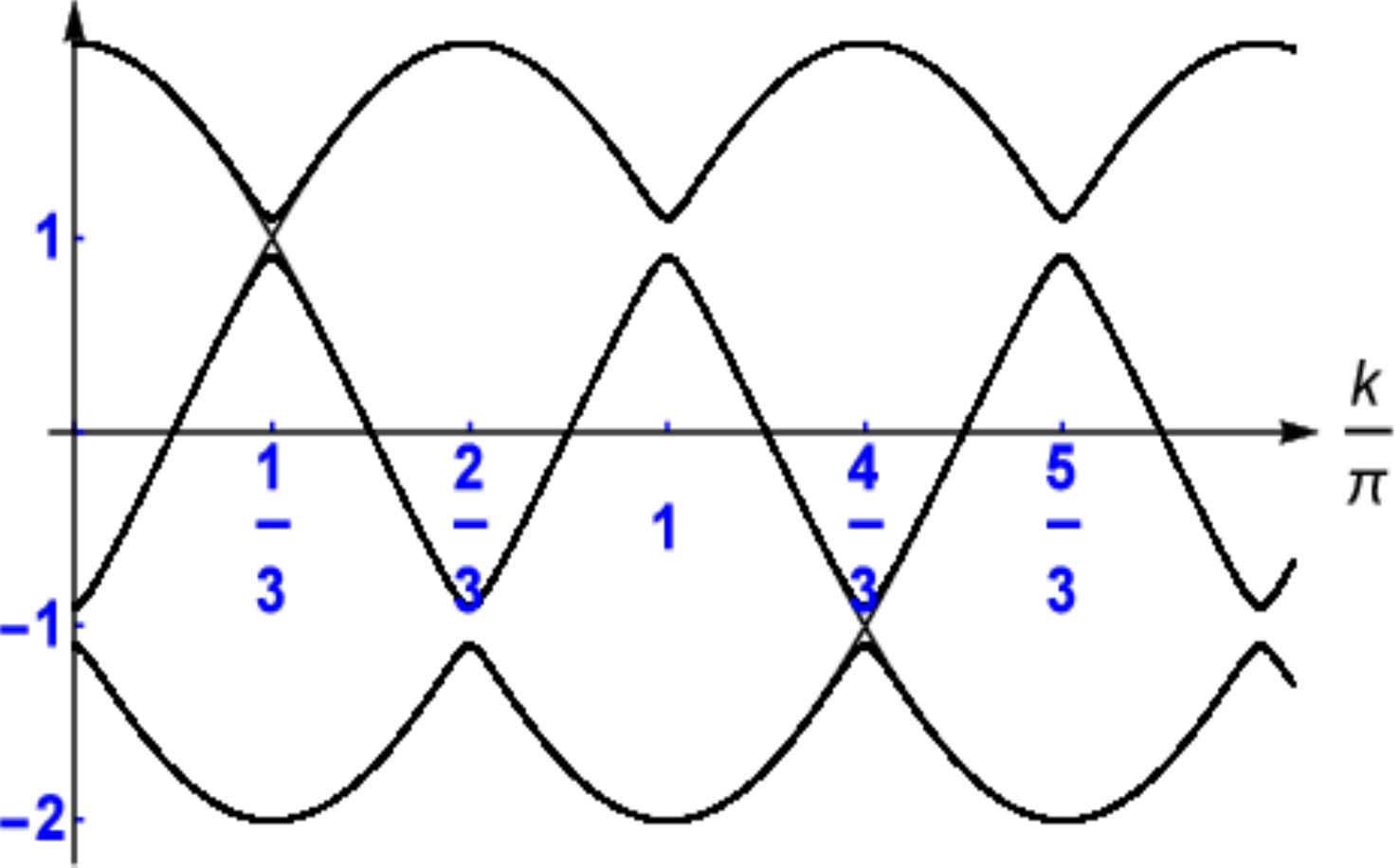} b)}
\end{minipage}
 \centering{\leavevmode}
    \begin{minipage}[h]{.75\linewidth}
\center{\includegraphics[width=\linewidth]{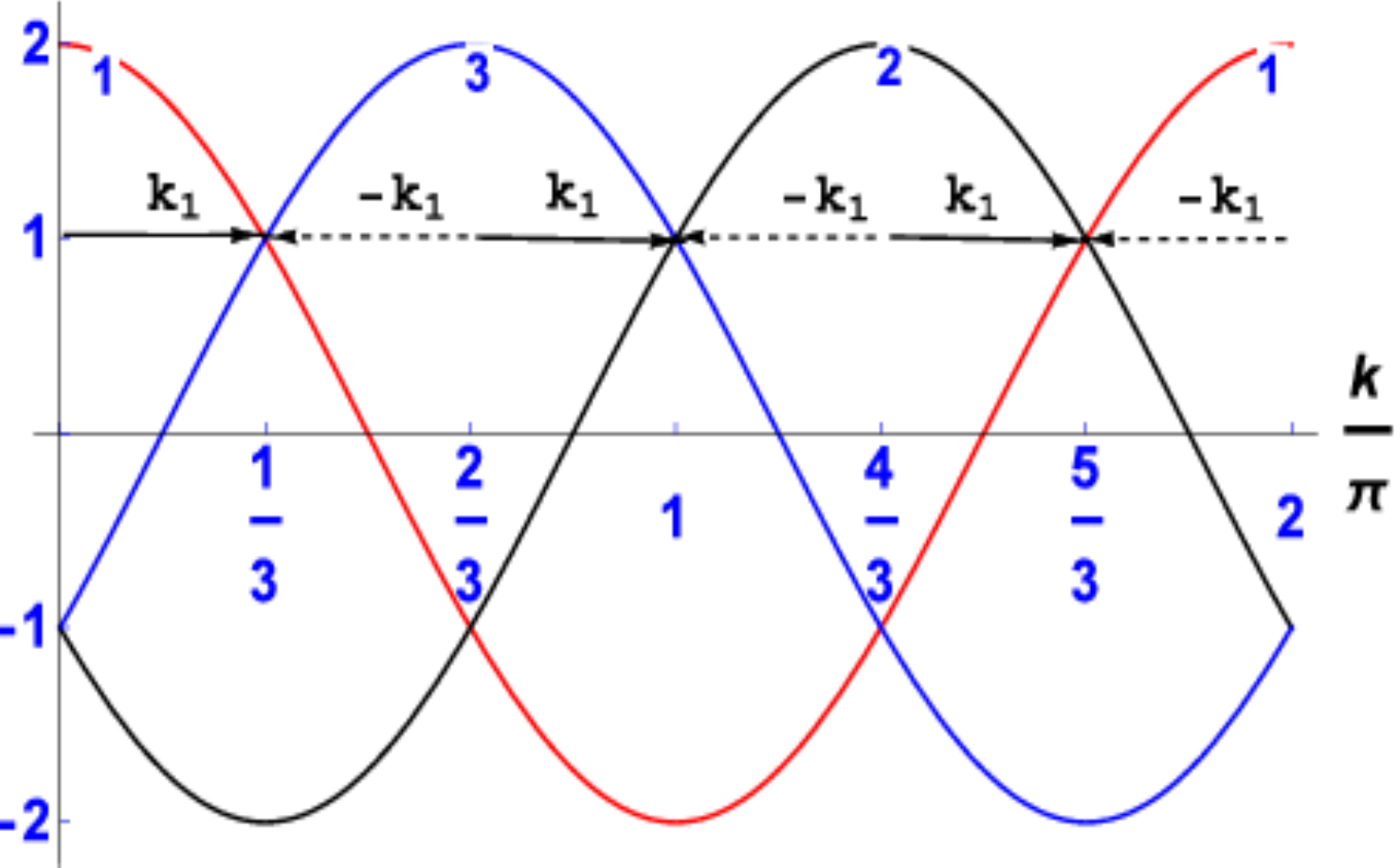} c)}
\end{minipage}
\caption{(Color online)
The band structure in the stripe geometry for $\phi=\frac{1}{3}$ $t=1$ a), $t=\frac{1}{10}$ b) and $t=0$ c) (colored lines indicate the numbered chains in a magnetic cell).  The topological structure of the spectrum does not change with the value $t>0$, the total number of the edge modes intersecting the gaps is conserved.
  }
\label{fig:1}
\end{figure}
The term ${\cal H}(\epsilon)$ determines the fermion states in the region of the energy $\epsilon$, the total Hamiltonian ${\cal H}(\epsilon)+{\cal H}(-\epsilon)$ does not break the $U(1)$ symmetry of the Hamiltonian (1). The Hamiltonians (2) define the  $x$-chains of isolated dimers, it was proposed by Kitaev \cite{K} for describing topological superconductivity in the chain of spinless fermions. The tunneling of fermions located at the nearest-neighbor $y-$chains opens the gaps $\Delta=\tau$ in the spectrum at the points $k_p$, $\pm 1$. Taking into account the numerical calculations of the spectrum, we can define value of $\tau$, assuming $ \tau = 2 t $. Considering a sample of a  cylindrical shape, where $N/q$ is a positive integer, we find that Majorana fermions located at the ends of the sample are free at $k_1=\frac{\pi}{3}$, zero energy edge states are realized at $k_1=\frac{\pi}{3}$ (see in {fig:1}a),b)).  The Majorana fermions are paired in the dimers with energy $\pm \frac{\tau}{2}$, the operators $\chi(1)$ and $\gamma(N)$  are free Majorana fermions at the energy equal to $1$, they remain unpaired and form zero energy edge states. In the weak t-limit, the velocities of the edge modes  at the middle of the gaps are $ \mp 2 \sin k_1 $; the intersection points $k_p$ are weakly depend on t. Free Majorana fermions states existing in the energy $ -1$ are associated with  Majorana operators $\gamma(1)$ and $\chi(N)$. The Majorana fermions have opposite chirality (see in {fig:1}), at energies $1$ and $ -1$, the Chern numbers have opposite signs -1 and 1, respectively.

Below we consider more general case of a rational flux $\phi =\frac{1}{6}$. Numerical calculations of the fermion spectrum are shown in {fig:2}. The topological structure of the subbands is determined by the following Chern numbers $\{1,1,-4,1,1\}$, here two (gapless) subbands in the center of the spectrum are one topological subband with the Chern number $-4$. The behavior of the edge modes in the first gaps, which correspond to the energies $\epsilon=\pm \sqrt3$ at $k_p=\frac{(2p-1)\pi}{6}$, ($p=1,...,6$) $t=0$, is described by the Hamiltonian (2) analogously to the flux $\phi =\frac{1}{3}$. The gaps are equal, $\Delta_1\simeq 2|t|(1-0(t^2))$
and $\tau \simeq 2 t$ at $t<<1$.
The generalization is based on the observation, that the Kitaev chains, with the next-nearest neighbor hoppings of fermions between different y-chains, can describe two zero energy states of Majorana fermions located at each boundaries.
In the $t\to 0$ limit the energies of the fermions in the y-chains are shifted in the phase $\frac{\pi}{3}$, the points  $k_p=\frac{\pi p}{3}$, $\epsilon=\pm 1$ ($p=1,...,6$) correspond to the intersections of the energies of fermions belonging to the next-nearest neighbor $y-$chains. At $\varepsilon=\pm 1$ the gaps are equal, $\Delta_2\simeq 2t^2(1-0(t^2)).$

The corresponding Hamiltonian ${\cal H}(\epsilon =1)+{\cal H}(\epsilon =-1) $ describing the energy states in the gaps ($\Delta_2$) is determined by two isolated chains, it is a generalization of (2)
\begin{eqnarray}
{\cal H}_{eff}(\epsilon =1) = i\frac{\tau}{2} \sum_{n=1}^{N-2} \chi_{n} \gamma_{n+2}+ i \frac{\tau}{2}\sum_{n=2}^{N-3} \chi_{n} \gamma_{n+2},\nonumber \\
{\cal H}_{eff}(\epsilon =-1) = i\frac{\tau}{2} \sum_{n=1}^{N-2}\chi_{n+2} \gamma_{n}+ i\frac{\tau}{2}\sum_{n=2}^{N-3} \chi_{n+2} \gamma_{n},
\label{eq:H2}
\end{eqnarray}
where $\tau\simeq 2 t^2$ at $t\ll 1$ is determined the hopping integral of fermions between the next-nearest neighbor $y$-chains. From numerical calculations it follows that the effective constant of tunneling of fermions located in the chains at a distance $\delta$ is  determined by  $\tau (\delta) \simeq \zeta t^\delta$,  the value of $\zeta$  is decreased from 2 to 1 with $\delta$.

The Hamiltonians (3) define the zero energy states of Majorana fermions located at the ends of the chains with even and odd lattice sites (see in {fig:2}).  The wave functions of Majorana fermions in the gaps $\Delta_1$ and $\Delta_2$ are shown in {fig:3}a) and {fig:3}b), the chiral gapless edge modes are localized at the boundaries of the sample. The edge modes are localized in the gaps, they populate the gaps and merge with the bulk states. The edge modes cross the gaps connecting the lower and upper subbands and intersect at $k_p$ due to Kramers degeneracy.

In the center of the spectrum the effective Hamiltonian has the form ${\cal H}_{eff}(+0)+{\cal H}_{eff}(-0)$, it determines the band states of fermions. Thus, the fermion states in the center of the spectrum are gapless ones for arbitrary value of a magnetic flux. Considering the structure of the edge modes for an arbitrary rational flux at $t<<1$ we obtain the diophantine equation \cite{D1,D2}, which is true for t=1,
the systems with different $t$ are topologically equivalent.

\begin{figure}[tp]
     \centering{\leavevmode}
    \begin{minipage}[h]{.85\linewidth}
\center{\includegraphics[width=\linewidth]{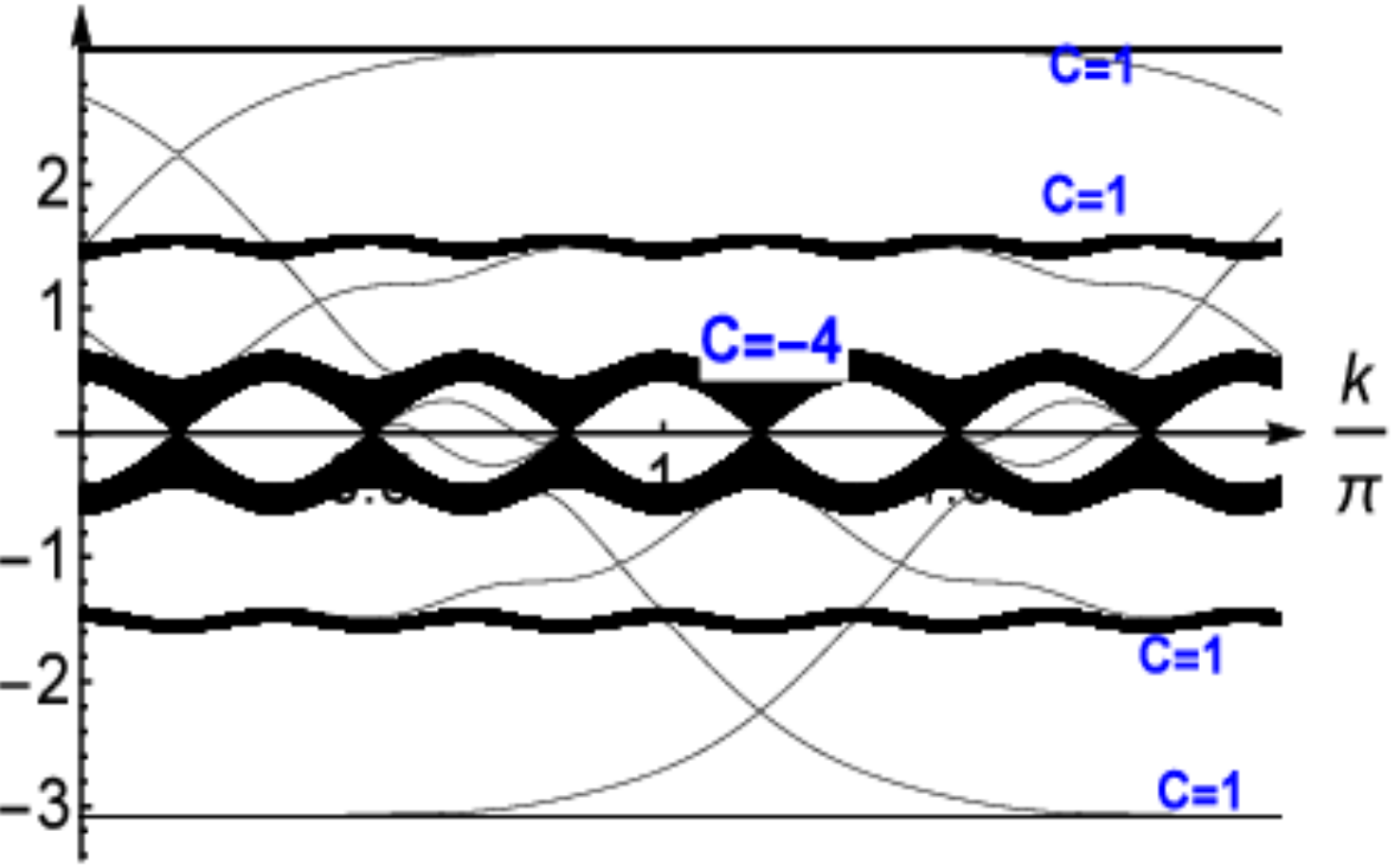} a)}
\end{minipage}
 \centering{\leavevmode}
    \begin{minipage}[h]{.85\linewidth}
\center{\includegraphics[width=\linewidth]{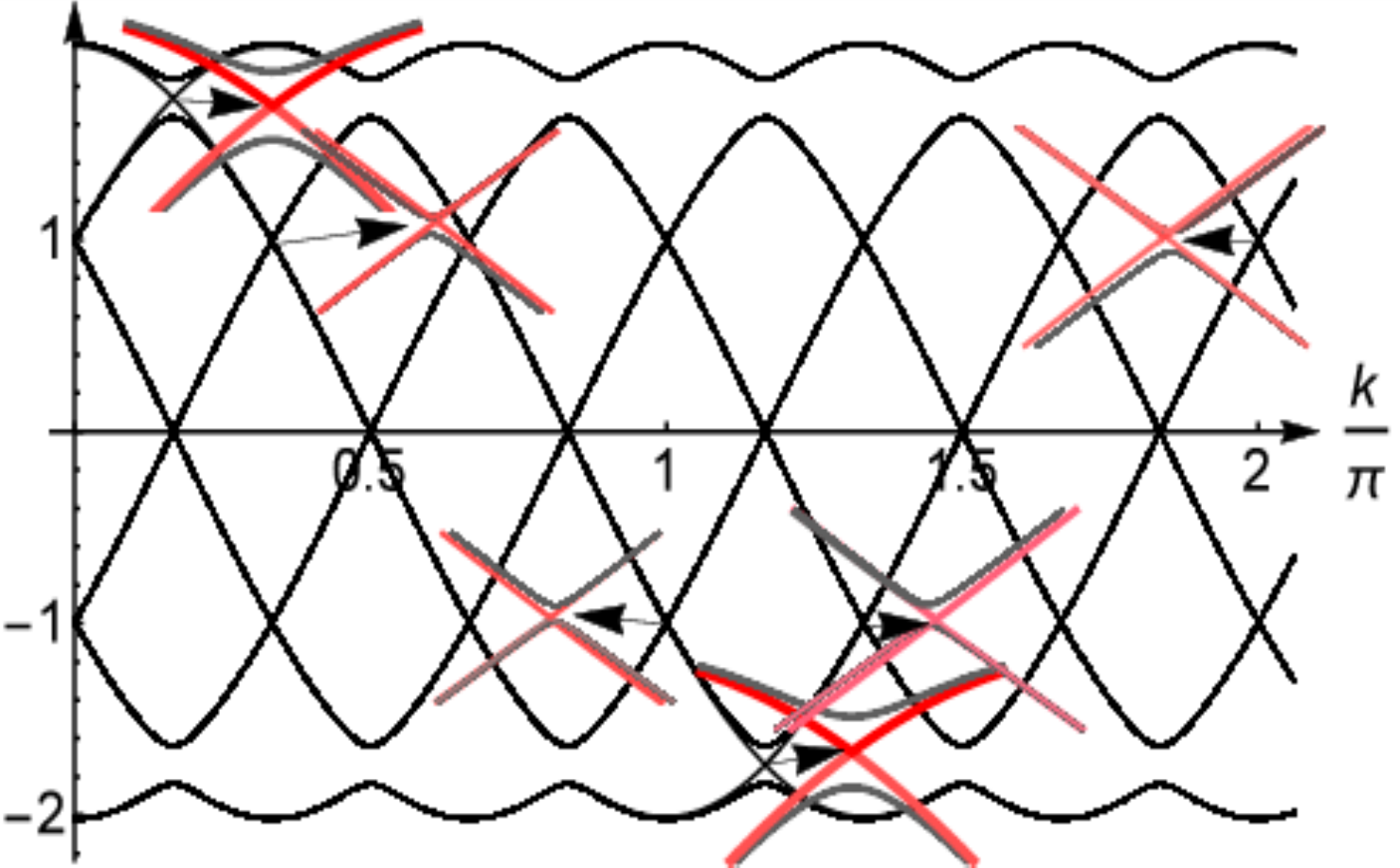} b)}
\end{minipage}
\caption{(Color online)
The band structure in the stripe geometry for $\phi=\frac{1}{6}$ strong $t=1$  a) and weak $t = \frac{1}{10}$ coupling b). The insets in b) zoom the regions with the edge modes (marked in red) around $\{ \frac{\pi}{6},\sqrt 3 \}$, $\{\frac{7\pi}{6}, -\sqrt 3 \}$, $\{ 2\pi, 1  \}$, $\{ \frac{\pi}{3}, 1 \}$, $\{ \pi, -1 \}$, $\{ \frac{4\pi}{3}, -1 \}$.
  }
\label{fig:2}
\end{figure}

\begin{figure}[tp]
     \centering{\leavevmode}
    \begin{minipage}[h]{.85\linewidth}
\center{\includegraphics[width=\linewidth]{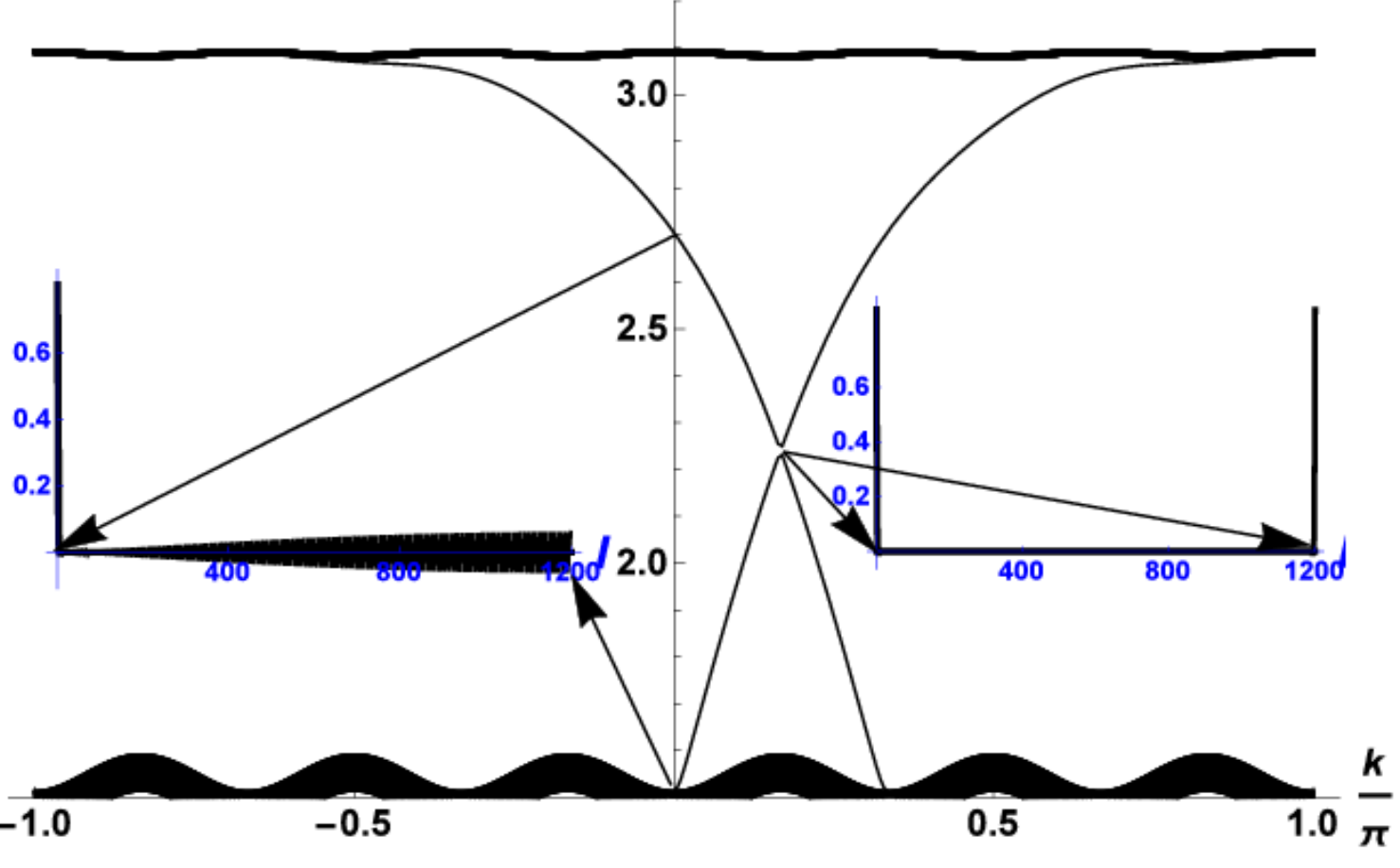} a)}
\end{minipage}
 \centering{\leavevmode}
    \begin{minipage}[h]{.85\linewidth}
\center{\includegraphics[width=\linewidth]{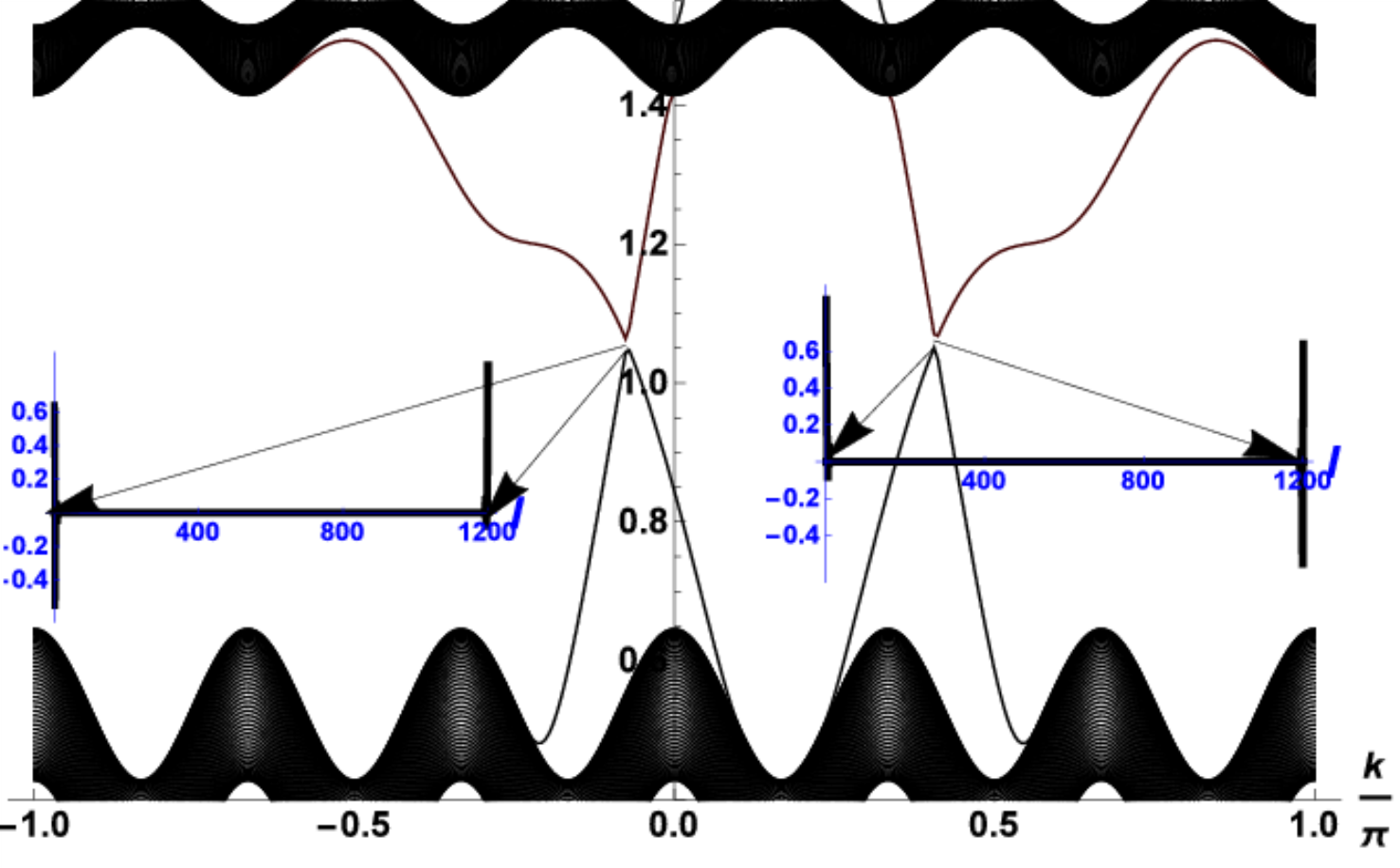} b)}
\end{minipage}
\caption{(Color online)
The spectrum of gapless Majorana fermions as a function of $k$ for $t=1$, $\phi=\frac{1}{6}$. The inserts exhibit a wave function of fermions (its real part, and imaginary part has the same behavior) defined at the lattice sites $l$ (along the x direction, N=1200).  The edge modes are shown in the gap $\Delta_1$ at $k=0,\epsilon=1.51$; $k=0,\epsilon=2.71$; $k=0.167 \pi, \epsilon=2.24$ a) and in the gap $\Delta_2$ at $k=0.405 \pi,\epsilon=1.05$; $k=-0.075\pi, \epsilon=1.054$ b).
  }
\label{fig:3}
\end{figure}

\subsection{An irrational flux $\phi =\frac{1}{\sqrt {8}}$}

\begin{figure}[tp]
     \centering{\leavevmode}
    \begin{minipage}[h]{.95\linewidth}
\center{\includegraphics[width=\linewidth]{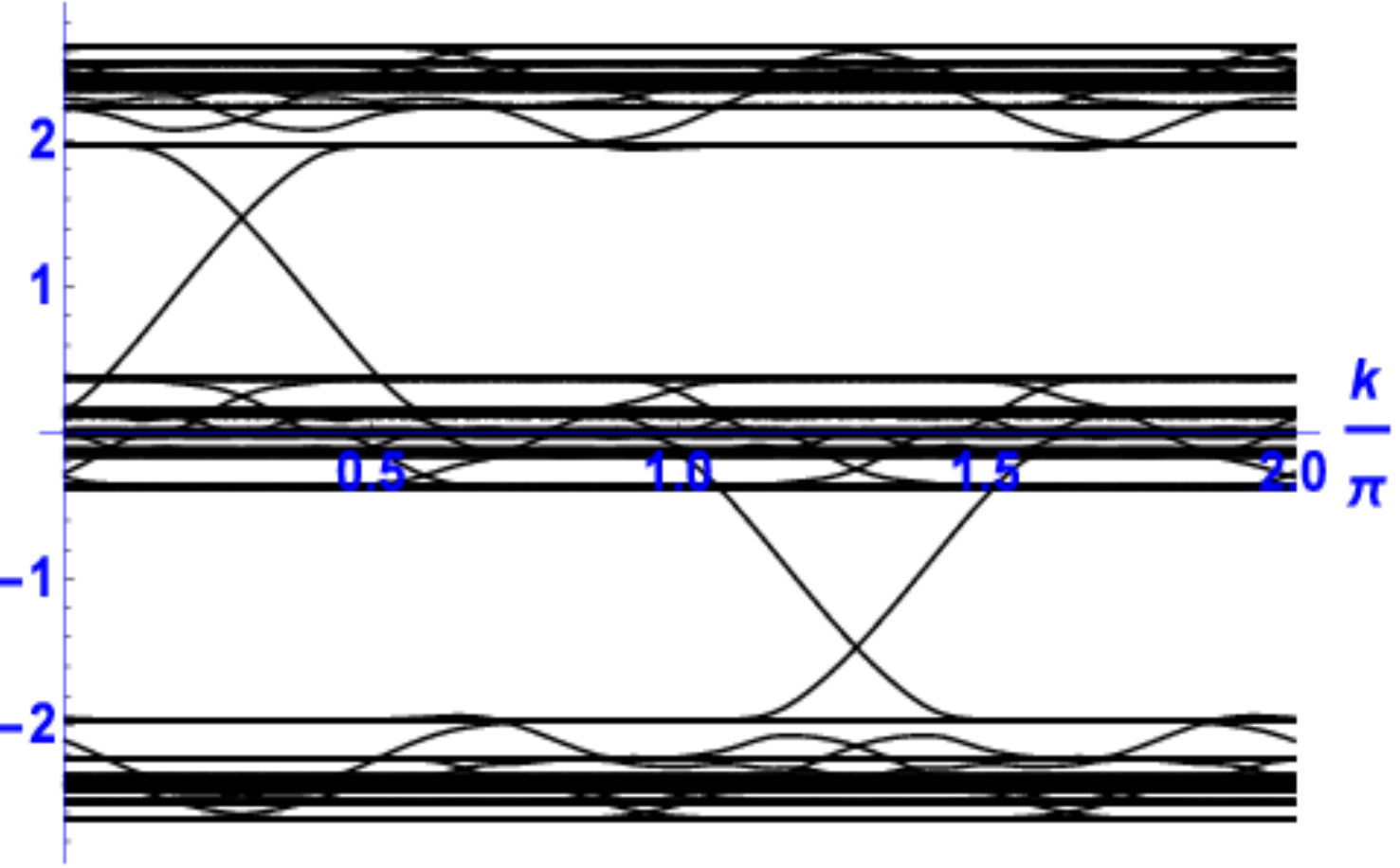} a)}
\end{minipage}
 \centering{\leavevmode}
    \begin{minipage}[h]{.95\linewidth}
\center{\includegraphics[width=\linewidth]{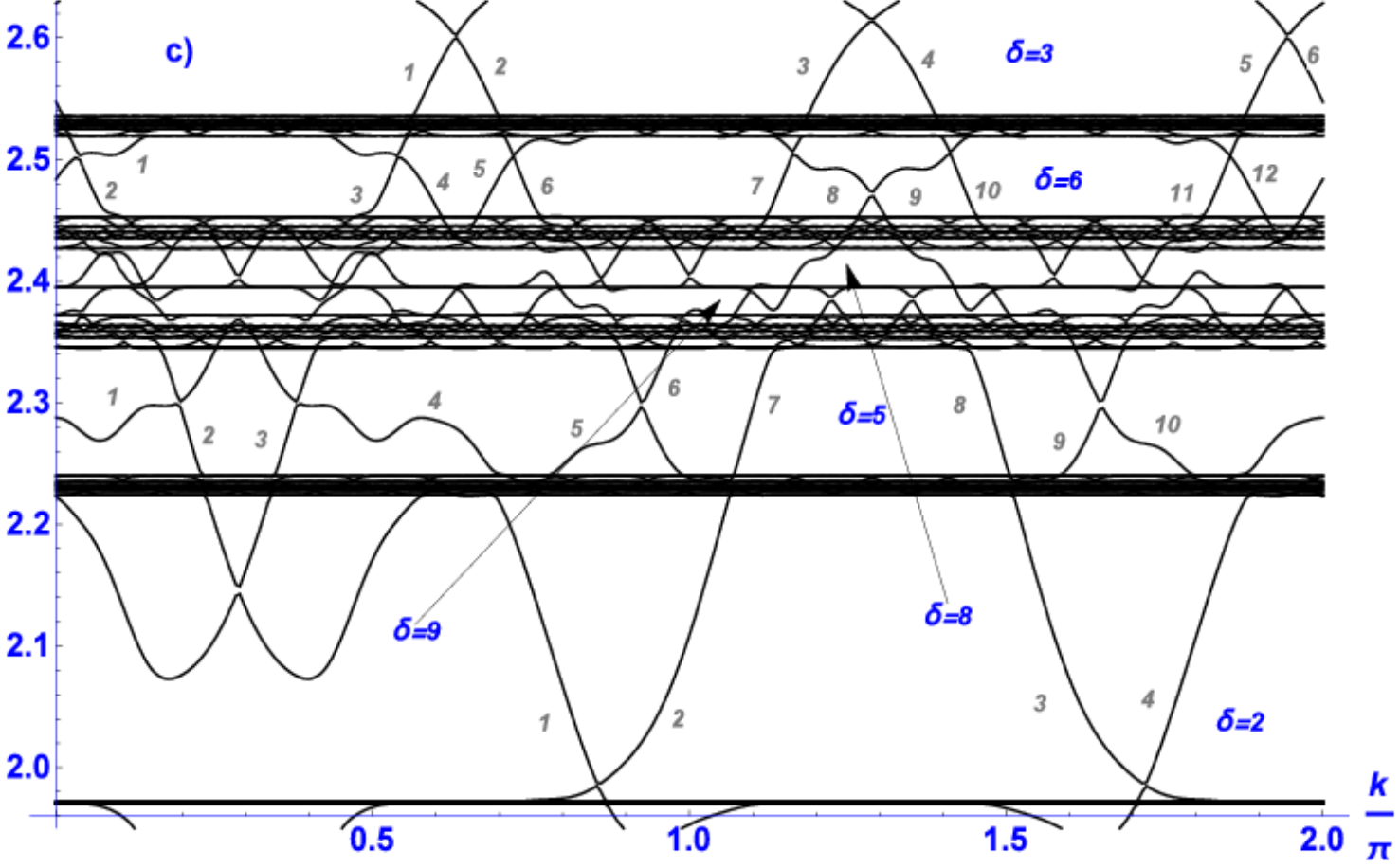} b)}
\end{minipage}
\caption{(Color online)
The band structure in the stripe geometry for $t=1$  $\phi=\frac{1}{\sqrt 8}$  (N=300) a), a fine structure of the high-energy subband  b). $2\delta$ denotes the total number of gapless edge modes in the gaps that separate the corresponding subbands of a fine structure of the spectrum (the modes are numerated).
  }
\label{fig:4}
\end{figure}

This approach makes it possible to study the topological structure of the fermion spectrum in the case of an irrational flux, the Hamiltonians (2),(3) describe the behavior of fermions for an arbitrary magnetic flux. For an irrational flux, a magnetic translational invariance of the system is broken, and the diophantine equation \cite{D1,D2} is not valid. As an example, we consider the irrational flux $ \phi = \frac {1} {\sqrt {8}}$ as transformation of the spectrum with the rational flux $\phi = \frac {1} {3}$ (see in  {fig:1}a). Numerical calculations of the excitation spectrum for a rational flux in the Hofstadter strips with open boundary conditions were obtained for samples, where $\frac{N}{q}$ is a positive integer. When calculating the spectrum for an irrational flux that is approximated by rational numbers $\phi \simeq \frac{p}{q}$, we assume that $N < q$. In this case numerical calculations correspond to the spectrum with an irrational flux to within $\frac{1}{N}$. Numerical calculations of the fermion spectrum for $\phi=\frac{1}{\sqrt 8}$, $t=1$, obtained on the sample with size $N=300$, are shown in {fig:4}a). In the case of the irrational flux a fine structure of the topological subbands is formed.
The behavior of gapless edge modes, associated with these subbands, depends on the size of the sample, the shape of these modes is different for different values of $N$, while their total number is conserved. The Chern number is determined by the integral of the Berry curvature over the Brillouin zone, reduces to integration along the Brillouin zone boundary. In the case of an irrational flux, the Chern number is not well defined for an indefinite unit cell \cite{Ch1}, but the total number of chiral gapless edge modes is conserved. Using the numerical calculations at t=1 and proposed approach at $t<<1$  we shall show  below, that in the case of an irrational flux, the total number of chiral edge mode in the gaps of the fermion spectrum is fixed and determines the Hall conductance. As was to be expected, the topological structure of the spectrum is not critical with respect to the value of t.

The fine structure of the subbands is calculated using the following approach. For example, consider the high-energy subband, which lies at $\varepsilon> 1 $ (see in {fig:4}a)). In the weak t-limit $ t \to 0 $, the gaps at the energies $ \varepsilon =\pm 2  \cos\frac {\pi\delta}{q} $ and $\varepsilon \neq 0$ divide this band into the subbands. $\delta$ determines the distance between $ y-$chains of spinless fermions that tunnel, forming gaps in the spectrum for these energies. For $ \varepsilon> 1 $ we have the following set: $ \delta = 2 \Rightarrow \varepsilon = 1.2114 $, $\delta= 5 \Rightarrow \varepsilon = 1.49091 $, $\delta=9 \Rightarrow  \varepsilon =1.68196$, $\delta = 8 \Rightarrow \varepsilon = 1.71643 $, $\delta= 6 \Rightarrow  \varepsilon = 1.85648 $, $\delta= 3 \Rightarrow \varepsilon = 1.96379 $ ....

We have considered all possible gaps in the fine structure of the high-energy subband
when $\delta$ changes from 1 to 10. The value of $\delta$ determines also the total number of chiral gapless edge modes, localized at a boundary, or the Hall conductance for a given occupied subbands in the fine structure of the high-energy band. The calculation of the fine structure of this band is shown in {fig:4}b) for $t=1$, the energies corresponding to the gaps vary, but their sequence is conserved, which makes it possible to reconstruct the topological structure of the spectrum for any $t <\sim 1$. In the case of irrational fluxes the systems with different $t$ are also topologically equivalent.

\section{The Hofstadter model of interacting electrons}

The nontrivial topological properties of the 2D systems are manifested in the existence of a nontrivial value of the Chern number and chiral gapless edge modes. Below, we consider in detail the stability of edge modes taking into account for the on-site Hubbard interaction between electrons. By introducing spin degrees of freedom, it is possible to add the on-site Hubbard interaction $U$ (U is positive for repulsive interaction) in the Hamiltonian (1)
\begin{eqnarray}
{\cal H} ={\cal H}_0 +U \sum_{n,j} \left(n_{n,j;\uparrow}-\frac{1}{2}\right) \left(n_{n,j;\downarrow}-\frac{1}{2}\right),
\label{eq:H3}
\end{eqnarray}
where $\sigma =\uparrow,\downarrow$ determines the spin of electron, the first term in (4) takes into account the fermions with different spins, $n_{l,j;\sigma}$ is the density operator.

In the limit $t \to 0$ the model (4) is reduced to the isolated Hubbard $y-$ chains. In the case of a weak on-site interaction $U << 1 $  the electron spectrum of the Hubbard chain \cite{KOR} is renormalized slightly. We study the behavior of the Hofstadter model of interacting electrons in the case $U\approx \tau <<1$ and $U, \tau >>\mu_B H$, where $ \mu_B $ is the Bohr magneton. In this approximation, the Zeeman energy in the Hamiltonian (4) can be ignored. We stress our consideration on the behavior of edge modes in the gaps for a rational flux for the energy $\varepsilon \to 0$ reckoned from the Fermi energy. The Fermi energy lies in the gap with a half-filled subband in which the gap appears due to tunneling of fermions between the chains.
As we noted above, only two states of spinless fermions into the gaps with fixed momenta $-k_0(n-1)\pm \pi\delta/q$, and energy $\varepsilon$, defined at one site $n$, tunnel between the $y-$chains, these states have different chirality with energy $\varepsilon$ (see in {fig:1}c)).

The density of fermions at the site $n,j$, the energy $\varepsilon$ is determined as $n_{n,j;\sigma}=n_{n;\sigma}(\varepsilon)=\frac{i}{2}\gamma_{n;\sigma} \chi_{n;\sigma}+\frac{1}{2}$.
Taking into account that $\sum_{j}( n_{n,j;\uparrow}-\frac{1}{2})(n_{n,j;\downarrow}-\frac{1}{2})=\frac{1}{N}\sum_{k} m (n,k;\uparrow) m(n,-k;\downarrow)$,  where $m(n,k;\sigma)=\sum_{j}(n_{n,j;\sigma}-\frac{1}{2})\exp (i k j)$, we obtain
$\frac{1}{N}\sum_{k} m(n,k_0(n-1);\uparrow)m(n,k_0(n-1);\downarrow)=
- \frac{1}{4}\gamma_{n;\uparrow}\chi_{n:\uparrow} \gamma_{n;\downarrow}\chi_{n:\downarrow}$,
here $m(n,k_0(n-1);\sigma)=n_{n;\sigma}(\varepsilon)-\frac{1}{2}$.
We used the $\frac{2\pi}{q}$-periodicity of the bulk spectrum with the period $k_0$.
An  effective low-energy Hamiltonian has the following form
\begin{eqnarray}
&&{\cal H}_{eff}(\varepsilon,\delta) = i\frac{\tau (\delta)}{2} \sum_{\sigma} \sum_{n=1}^{N-\delta}\gamma_{n;\sigma} \chi_{n+\delta;\sigma} \nonumber \\
&&-\frac{U}{4}\sum_{n}  \gamma_{n;\uparrow} \chi_{n;\uparrow}\gamma_{n;\downarrow} \chi_{n;\downarrow},
\label{eq:H4}
\end{eqnarray}
where $\delta$ denotes also the number of the chains of electrons with the hopping integral $\tau(\delta)$ between electrons located at the sites on the distance $\delta$, Majorana fermions are defined for fermions with different spins.

Taking into account the spin freedom of fermions, the Chern number for electron subbands and the gapless edge modes  doubles. As a result, at $U=0$ and $\delta=1$ the Hamiltonian  (5) determines the two zero energy edge modes on each boundary which correspond to fermions with different spin.
The Hamiltonian (5) with the on-site Hubbard interaction is mapped to a noninteracting fermion model, which can be diagonalized exactly  \cite{MN}. The model is reduced to the Kitaev chain of spinless fermions (2) with the hopping integral $\tau (\delta)$ and the chemical potential $U$  \cite{MN}. Indeed, the ground state energy and the energy of the excited states are calculated exactly. For arbitrary $\tau(\delta)$ and $U$, the energy spectrum of the chain (5) is gapped, having the form
\begin{equation}
\epsilon_{k_x} = \pm \frac{1}{2}\sqrt{\left(\frac{U}{2}-2\tau(\delta)\right)^2+4 U \tau(\delta)\sin ^2 \frac{\delta k_x}{2}}.
\label{eq:H5}
\end{equation}
For $U = 4\tau(\delta),-4\tau(\delta)$ the gap vanishes, in this case we have for $ k_x \to 0, \frac{\pi}{\delta}$ a spectrum (6) linear in $ k_x$. According to \cite{MN},
the ground state degeneracy is dependent on whether $\xi=\frac{U}{\tau(\delta)}>4$ or otherwise.
Interestingly, for $|\xi|<4$ the equation for wave vectors (which follows from the boundary conditions) has one real root less than in the previous case along with an imaginary root associated with the ends of the chain.
In the thermodynamic limit there are zero energy Majorana states. These zero mode Majorana states are absent for $|\xi| >4$ and its presence for $|\xi|<4$ increases the degeneracy by two, since excitation of the Majorana mode  does not change the energy of the system.  The low energy spectrum is shown in {fig:5} as a function of $\xi$. The zero energy edge states are realized in the interval $[-4,4]$.
The number of the chiral gapless edge modes follows the Chern number, so in the thermodynamic limit the system has a quantum phase transition at $|\xi |=4$. The phase transition is realized between the phase states with different Chern numbers. The value $\tau (\delta)$ decreases with increasing of $\delta$ or decreasing $|\epsilon|$, so the phase states with the Chern numbers, for which $ |\frac{U}{\tau(\delta)}|>4$, are topological trivial states when the interaction is taking into account. The topological ambitions of the low energy subbands are limited by a weak coupling $U<4\tau (\delta)$. In this case, the topological phase transition is realized also when the filling is changed. The criterium of realization of the topological state at the $\gamma$-filled band separated by a gap $\Delta_\gamma\simeq\tau(\delta)$ is defined as $\Delta_\gamma >\frac{U}{4}$.

\begin{figure}[tp]
    \centering{\leavevmode}
     \centering{\leavevmode}
 \begin{minipage}[h]{.75\linewidth}
\center{\includegraphics[width=\linewidth]{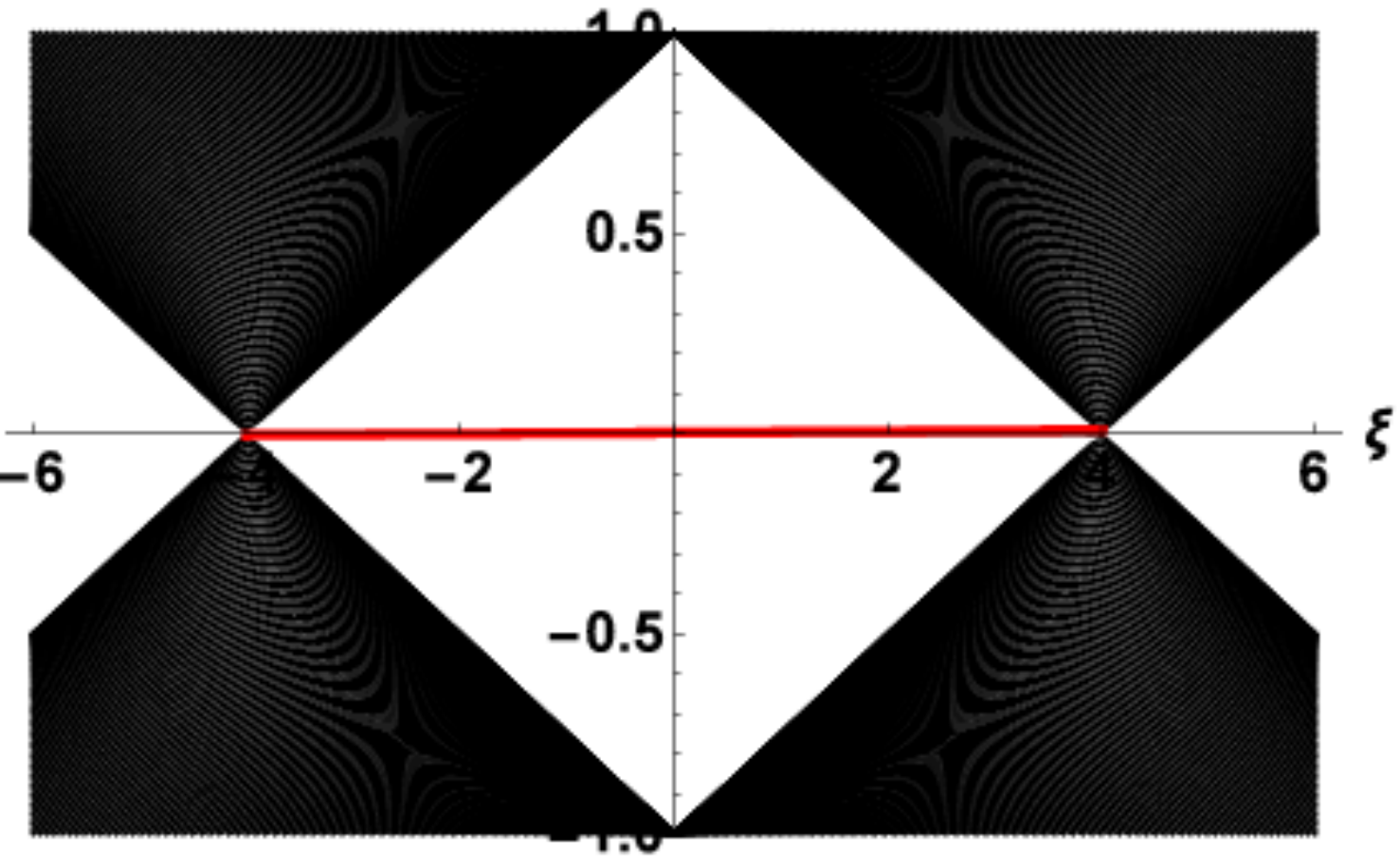}}
\end{minipage}
\caption{(Color online)
Low energy spectrum of the chain (5) with open boundary conditions as function of $\xi=\frac{U}{\tau (\delta)}$. The zero energy edge modes are marked in red in the interval [-4,4].
 }
\label{fig:5}
\end{figure}

\section{Conclusion}

We have studied the behavior of 2D fermions in the Hofstadter model with a rational magnetic flux through an unit cell, focusing on the realization of the chiral gapless edge modes. The fine structure of the bands is characterizes by the Chern numbers, which are well defined for insulator phase. The chiral gapless edge modes are described in the framework of the Kitaev chain with effective hopping integral of Majorana fermions.
In the case of an irrational flux the topological properties of fermions is not described in the framework of the Chern insulator, in this case the Chern number is not clearly defined.  The topology of the fine structure of the subbands is determined by the total number of chiral edge modes that determine the Hall conductance.

The 2D topological insulators that support chiral gapless edge states are extremely susceptible to the on-site Hubbard interaction. Strong interaction generically destroys the Majorana edge states,
in the case of weak interaction, a regime, in which the Majorana edge states persist, is realized. The stability of the chiral edge modes and, consequently, the phase of topological insulator is determined by the following relation between the values of gap $\Delta$ and the on-site Hubbard interaction $U <4 \Delta$.
At the point $U=4\Delta$ a topological phase transition occurs in which the Chern integers of two subbands, which degenerate as the gap closes, change by integer amounts.
The subbands near the centrum of the fermion spectrum with smaller values of gaps are topological trivial due to the presence of the interaction between fermions. We find that moderate on-site Hubbard interaction did not affect the phase diagram qualitatively but led to nontrivial quantitative changes in the phase boundary.
Obtained results determine the stability of the topological state against interactions, they can be applicable to different 2D topological insulators.

This research was supported by the budget program 6541230 "Support for the development of priority areas of research".
\begin{center}
\end{center}

\end{document}